# Raman Evidence for Superconducting Gap and Spin-Phonon Coupling in Superconductor $Ca(Fe_{0.95}Co_{0.05})_2As_2$

Pradeep Kumar[1], Achintya Bera[1], D. V. S. Muthu[1], Anil Kumar[2], U. V. Waghmare[2],

L. Harnagea[3], C. Hess[3], S. Wurmehl[3], S. Singh[3], B. Büchner[3] and A. K. Sood[1,*]

[1]Department of Physics, Indian Institute of Science, Bangalore -560012, India

[2]Theoretical Sciences Unit, Jawaharlal Nehru Centre for Advanced Scientific Research, Bangalore -560064, India

[3] Institute for Solid State Research, IFW Dresden, D-01171 Dresden, Germany

**ABSTRACT**

Inelastic light scattering studies on single crystal of electron-doped $Ca(Fe_{0.95}Co_{0.05})_2As_2$ superconductor, covering the tetragonal to orthorhombic structural transition as well as magnetic transition at $T_{SM}$ ~ 140 K and superconducting transition temperature $T_c$ ~ 23 K, reveal evidence for superconductivity-induced phonon renormalization; in particular the phonon mode near 260 cm$^{-1}$ shows hardening below $T_c$, signaling its coupling with the superconducting gap. All the three Raman active phonon modes show anomalous temperature dependence between room temperature and $T_c$ i.e phonon frequency decreases with lowering temperature. Further, frequency of one of the modes shows a sudden change in temperature dependence at $T_{SM}$. Using first-principles density functional theory-based calculations, we show that the low temperature phase ($T_c < T < T_{SM}$) exhibits short-ranged stripe anti-ferromagnetic ordering, and estimate the spin-phonon couplings that are responsible for these phonon anomalies.

## 1. INTRODUCTION

The discovery of superconductivity in the iron based superconductors $RFeAsO_{1-x}F_x$ (R = La, Sm, Ce, Nd, Pr and Gd ) and $AFe_{2-x}Co_xAs_2$ (A = Ca, Sr and Ba) [1-6] with transition temperature up to 55K has generated enormous interest to investigate these materials both theoretically and experimentally. The parent compounds LaFeAsO and $AFe_2As_2$ (termed as so called "1111" and "122" systems, respectively) exhibit long range antiferromagnetic (AFM) order akin to cuprates which is suppressed on doping [7-8]. There are a numbers of different scenarios proposed for the possible superconducting pairing mechanism in these iron pnictides [9-13], including the importance of AFM spin fluctuation and electron-phonon coupling through spin-channels. Recently, inverse iron isotope effect has been observed in 122 systems, suggesting an exotic coupling mechanism in these superconductors involving phonons [13].

One of the most important quantities for understanding the pairing mechanism in high temperature superconductors is the superconducting gap ($2\Delta$), whose magnitude and structure are intimately linked to the pairing mechanism. Raman spectroscopy has been proved to be a powerful technique to investigate the superconductivity-induced phonon renormalization in high $T_c$ superconductors as well as phonon anomalies well above $T_c$ [14-19]. A few temperature-dependent Raman studies have been reported on these iron based superconductors for "1111" [18, 20-21], "122" [22-26] and "11" ($FeSe_{1-x}$) [27] systems. In case of $Sr_{1-x}K_xFe_2As_2$ [22] no phonon anomaly was seen as a function of temperature. However, in another study of $R_{1-x}K_xFe_2As_2$ (R= Ba, Sr) [24] the linewidths of the phonon modes involving Fe and As near 185 cm$^{-1}$ ( $A_{1g}$ ) and 210 cm$^{-1}$ ( $B_{1g}$ ) show a decrease below the spin-density-wave transition temperature $T_s$ ~ 150 K, attributed to spin-density gap opening. Also, the frequency of the 185 cm$^{-1}$ mode shows a discontinuous change at $T_s$, signaling first order structural transition accompanying the spin-density-wave transition at $T_s$. Similar results for the $B_{1g}$ mode ( near 210 cm$^{-1}$) are seen for Sr$_{0.85}$K$_{0.15}$Fe$_2$As$_2$ and Ba$_{0.72}$K$_{0.28}$Fe$_2$As$_2$ ($T_s$ ~ 140 K) [26]. In the parent compound CaFe$_2$As$_2$, the $B_{1g}$ phonon frequency ( 210 cm$^{-1}$) shows a discontinuous decrease at $T_s$ ~ 173 K and the $A_{1g}$ phonon ( near 190 cm$^{-1}$ ) intensity is zero above $T_s$, attributed to the first order structural phase transition and a drastic change of charge distribution within the FeAs plane [23]. The $E_g$ (Fe, As) phonon (~ 135 cm$^{-1}$ ) in Ba(Fe$_{1-x}$Co$_x$)$_2$As$_2$ ( x < 0.06) splits into two modes near the structural transition temperature ( $T_s$ ~ 100 to 130 K) linked to strong spin-phonon coupling [25]. Similarly, $E_g$ ( ~ 100 cm$^{-1}$) mode in case of FeSe$_{0.82}$ associated with the Se vibrations in *ab* plane shows anomalous hardening below $T_s$ attributed to the strong spin-phonon coupling [27-28].

However, there are no detailed temperature-dependent Raman studies on electron-doped CaFe$_2$As$_2$. In this paper we report such a study of single crystals of Ca(Fe$_{0.95}$Co$_{0.05}$)$_2$As$_2$ with $T_c$ ~ 23 K [29-30] in the temperature range of 4 K to 300 K covering the spectral range from 120 cm$^{-1}$ to 800 cm$^{-1}$. There are two motivating factors behind this work: first, Raman study of phonons can provide information on the superconducting state through superconductivity induced phonon renormalization. Second, influence on the phonon self-energy by other quasi-particles excitations, e.g spin-waves, can be probed as highlighted in the recent experimental and theoretical studies of these materials. Here, from temperature-dependent Raman scattering in tetragonal Ca(Fe$_{0.95}$Co$_{0.05}$)$_2$As$_2$, we present two significant results: (i) we find a strong evidence for the superconductivity-induced phonon renormalization, in particular a phonon mode near 260 cm$^{-1}$ shows anomalous hardening below $T_c$ attributed to an opening of the superconducting gap; (ii) we find evidence for the spin-phonon coupling below the magnetic transition temperature accompanied by the tetragonal-to-orthorhombic structural transition in the doped superconducting system.

## 2. METHODS

### 2.1 Experimental Details

Single crystals of nominal composition $Ca(Fe_{0.95}Co_{0.05})_2As_2$ with a superconducting transition temperature $T_c \sim 23$ K were prepared and characterized as described in ref. 29-30. Unpolarised micro-Raman measurements (with spectrum resolution $\sim 4$ cm$^{-1}$) were performed in backscattering geometry, using 514.5 nm line of an Ar-ion Laser (Coherent Innova 300) and Raman spectrometer (DILOR XY) coupled to a liquid nitrogen cooled CCD detector. The crystal surface facing the incident radiation is *ab* plane. Temperature variation was taken from 4 K to 300 K, with a temperature accuracy of ± 0.1K, using continuous flow He cryostat (Oxford Instrument).

### 2.2 Computational Details

Our first-principles calculations are based on density functional theory as implemented in the PWSCF [31] package. We use optimized norm-conserving pseudopotential [32] for Ca, As and ultrasoft pseudopotentials [33] for Co and Fe to describe the interaction between ionic cores and valence electrons, and a local density approximation (LDA) of the exchange energy functional. We use plane wave basis with a kinetic energy cutoff of 40 Ry in representation of wavefunctions and a cutoff of 240 Ry in representation of the charge density. We sampled integration over the Brillouin zone (of single unit cell) with 12x12x8 Monkhorst Pack Mesh [34]. Structural optimizations of $CaFe_2As_2$ and Ca(Fe$_{1-x}$Co$_x$)$_2$As$_2$ are done with $\sqrt{2}$ x $\sqrt{2}$ x1 supercells by minimizing the total energy using Hellman-Feynman forces and the Broyden-Flecher-Goldfarb-Shanno based method. Zone center (q = 0, 0, 0) phonon spectra are determined using a frozen phonon method (with atomic displacements of ± 0.04 Å) for the relaxed structure obtained at experimental values of the lattice constants. Self-consistent solution with different magnetic ordering, particularly ferromagnetic (FM) and G-antiferromagnetic types (AFM1), was rather demanding and was achieved with a mixing of charge density based on local density dependent Thomas-Fermi screening and a mixing factor of 0.1 (density from the new iteration with weight of 0.1). To facilitate comparison between theory and experiment within the accuracy of calculational framework and estimate the errors associated with use of pseudopotentials, we repeated all calculations for *x=0* with projector augmented wave potentials as implemented in a plane-wave package VASP [35-36]. Zone centre phonons were determined using frozen phonon method for structures obtained after internal relaxation for each of the nonmagnetic (NM), FM, AFM1 and stripe anti-ferromagnetic (AFM2) orderings.

## 3. RESULTS AND DISCUSSIONS

### 3.1. Raman Scattering from Phonons

CaFe$_2$As$_2$ has a layered structure belonging to the tetragonal *I4/mmm* space group. There are four Raman active modes belonging to the irreducible representation A$_{1g}$ (As) + B$_{1g}$ (Fe) +2E$_g$ (As and Fe) [23]. Figure 1 shows Raman spectrum at 4 K, revealing 3 modes labeled as S1 (205 cm$^{-1}$), S2 (215 cm$^{-1}$) and S3 (267 cm$^{-1}$). Spectra are fitted to a sum of Lorentzian functions. The individual modes are shown by thin lines and the resultant fit by thick line. Before we discuss assignment of modes S1 to S3 as phonon modes, we review the assignment of the Raman modes calculated and experimentally observed so far in "122" systems [22-26, 37]. In $Sr_{1-x}K_xFe_2As_2$ (x = 0, 0.4), the four Raman active modes have been observed and assigned as 114 cm$^{-1}$ ($E_g$ : As and Fe), 182 cm$^{-1}$ ($A_{1g}$ : As), 204 cm$^{-1}$ ($B_{1g}$ : Fe) and 264 cm$^{-1}$ ($E_g$ : As and Fe) [22]. The two modes observed in $CaFe_2As_2$ [23] are 189 cm$^{-1}$ ($A_{1g}$ : As) and 211 cm$^{-1}$ ($B_{1g}$ : Fe). In $Ba(Fe_{1-x}Co_x)_2As_2$ system, three observed modes have been assigned as 124 cm$^{-1}$ ($E_g$ : As and Fe), 209 cm$^{-1}$ ($B_{1g}$ : Fe) and 264 cm$^{-1}$ ($E_g$ : As and Fe) [25]. In $R_{1-x}K_xFe_2As_2$ (R = Ba, Sr), four Raman active modes have been identified as 117 cm$^{-1}$ ($E_g$ : As and Fe), 189 cm$^{-1}$ ($A_{1g}$ : As), 206 cm$^{-1}$ ($B_{1g}$ : Fe) and 268 cm$^{-1}$ ($E_g$ : As and Fe) [24]. However in another Raman study of $R_{1-x}K_xFe_2As_2$ (R = Ba, Sr) only one mode was observed and assigned as 210 cm$^{-1}$ ($B_{1g}$ : Fe) [26]. In a theoretical calculation by Hou et al [37] for $SrFe_2As_2$ the four calculated Raman modes in non-magnetic state are 138.9 cm$^{-1}$ ($E_g$ : As and Fe); 207.6 cm$^{-1}$ ($A_{1g}$ : As); 219.5 cm$^{-1}$ ($B_{1g}$ : Fe) and 301.2 cm$^{-1}$ ($E_g$ : As and Fe). The phonon frequencies for $BaFe_2As_2$ are close to the values for $SrFe_2As_2$. Keeping these reports in view and our density functional calculations (see Table II, III and fig. 2), we assign the modes S1 to S3 as S1: 205 cm$^{-1}$ ($B_{1g}$ : Fe), S2 : 215 cm$^{-1}$ ($A_{1g}$ : As) and S3: 267 cm$^{-1}$ ($E_g$ : Fe and As). We note from figure 1 that the linewidth of the mode S3 is larger than that of S1 and S2 modes. It is likely that the large linewidth of S3 (with eigen vectors in *a-b* plane) may arise from the disorder caused by the slight rotation of *a-b* plane in these layered crystals.

### 3.2. Temperature Dependence of the Phonon Frequencies

Figure 3 shows the peak frequency and linewidths of the three phonon modes as a function of temperature. The solid lines are linear fits to the data in a given temperature window. The following observations can be made: (i) The temperature dependence of the frequency of mode S1 shows a discontinuous change at $T_{SM}$. The frequency of the mode has anomalous temperature dependence below $T_{SM}$ (i.e the frequency decreases with lowering of the temperature). (ii) The temperature dependence of the mode S2 is anomalous in the entire temperature range of 4 K to 300 K. (iii) The frequency of the mode S3 also shows anomalous temperature dependence between $T_c$ and 300 K. Below $T_c$, the mode hardens on decreasing the temperature. (iv) The linewidths of mode S1 and S2 remains nearly constant with temperature, but on the other hand, the linewidth of mode S3 shows non-monotonic dependence on temperature below $T_{SM}$.

In superconductors, the opening of the superconducting gap (2Δ) below $T_c$ redistributes the electronic states in the neighborhood of the Fermi-surface and hence can change the phonon self-energy as seen in cuprates and other high temperature superconductors [14-19]. According to Zeyer et al [38] a change in phonon self-energy below $T_c$ is linked with the interaction of the phonons with the superconducting quasi-particles. Qualitatively, based on mode repulsion coupled excitation model, phonons above the 2Δ show hardening below $T_c$ whereas the phonons with frequency below the gap value can soften [38]. The anomalous hardening of mode S3 below $T_c$ (shown in Fig. 3) indicates a coupling of this phonon of $E_g$ symmetry involving the vibrations of Fe and As atoms to the electronic system. A microscopic understanding of this coupling may help in understanding the symmetry of the superconducting gap. Taking the phonon frequency of S3 mode as an upper limit of 2Δ, an estimate of $2Δ/k_BT_c$ is ~ 15, pointing to strong-coupling nature of superconductor. We note that in "122" systems experimental evidences of single and multiple gaps have already been reported from infrared spectroscopy [39], angle resolved spectroscopy [40-41], nuclear magnetic resonance [42-43] and muon spin rotation [44]. The reported gap values show a large variation from $2\Delta/k_BT_c$ ~ 1.6 to 10 in both electron-doped [39, 41, 44] and hole-doped systems [40, 42-43]. The origin of this large spread is still not understood.

The anomalous softening of the modes with decreasing temperature for all the three modes might be attributed to strong spin-phonon coupling, in line with other studies for similar pnictide systems. A recent study on the isotope effect in iron-pnictide '122' superconductors [45] suggests that electron-phonon interaction do play a role in the superconducting pairing mechanism via strong spin-phonon coupling. The coupling between phonons and spin degrees of freedom can arise either due to modulation of exchange integral by phonon amplitude [46-47] and/or by involving change in the Fermi surface by spin-waves provided phonon couples to that part of the Fermi surface [17]. Earlier Raman [24-25, 27] and neutron scattering studies [48-49] in iron-pnictides have also indicated strong spin-phonon coupling and have been supported by earlier theoretical calculation [12, 28, and 37]. In order to further elucidate the importance of strong spin-phonon coupling also in the present case of $Ca(Fe_{0.95}Co_{0.05})_2As_2$, we performed detailed DFT calculations.

**3.3. Theoretical Calculations**

We now present results of density functional theory based calculations to understand the role of magnetic ordering on phonons. To find out which magnetic ordering is relevant in the ground state at low temperatures, we carried out self-consistent total energy calculations for non-magnetic (NM) as well as different magnetic orderings i.e. FM, AFM1 and AFM2. Within Local spin density approximation (LSDA), our calculations initialized with FM, AFM1 and AFM2 type antiferromagnetic orderings relax to a nonmagnetic structure at the self-consistency. However, local stability of these magnetically ordered states could be achieved through use of an onsite correlation (Hubbard) parameter U ( = 4eV for Fe atoms) in the LSDA+U formalism. Calculated total energies of different magnetic orderings within LSDA+U description show that the stripe antiferromagnetic ordering has lowest energy (see Table I) and it should be prominent at low temperatures. We note a weak shear stress $\sigma_{xy}$ appearing in the stripe phase (Table I), which should give rise to a small orthorhombic strain as the secondary order parameter below $T_{SM}$.

To estimate the strength of spin-phonon coupling, if any, we determined phonon frequencies at wave-vector q = (0, 0, 0) using frozen phonon method for both the NM and stripe antiferromagnetic ordering (AFM2). Effects of spin-ordering on phonons are reflected in the frequencies of these two phases (see Table II). We find that the frequencies change by a large amount from NM phase to AFM2 phase (changes are of the order of 20-40 cm$^{-1}$), indicating the presence of a strong spin-phonon coupling in pure as well as doped CaFe$_2$As$_2$ systems. Calculated Raman active phonon modes are in reasonable agreement with the observed values (see Table II). We find that frequencies of the three Raman active modes estimated for a state with AFM2 ordering are smaller than those estimated for the NM ordering at both the compositions x = 0 and x = 0.25 studied here (see Table II). To estimate the spin-phonon couplings relevant to observed Raman spectra, we determined zone centre phonons for relaxed structures (kept at the experimental lattice constant) with AFM1, AFM2, FM and NM ordering (see Table III). It is evident from the frequencies that the spin-phonon coupling for all the three modes is strong. To understand the temperature dependence of phonon frequencies arising from the spin-phonon coupling, ($\lambda^{(1)}u^2*S_iS_j$, u being phonon coordinate and $S_i$ the spin on *i*th Fe) we construct a simple Ising spin-Hamiltonian [28] $H = J_1 \sum_{NN} S_i.S_j + J_2 \sum_{2^{nd} NN} S_i.S_j$, where $J_1$ and $J_2$ are the nearest and next nearest neighbor exchange interaction parameters. We note that though Heisenberg-type model should be used to capture the spin dynamics of complex systems such as *Fe*-based superconductors, it has been shown theoretically [50] that ground states of these systems have collinear stripe type anti-ferromagnetic spin ordering and for such systems Heisenberg and Ising models should give similar results. We use Monte Carlo (MC) simulations to obtain $T$-dependent spin-ordering. $J_1$ and $J_2$ are estimated from the energies of states with different magnetic ordering (see Table I) of CaFe$_2$As$_2$: $J_1 = -9$ meV and $J_2 = 19$ meV. The temperature dependence of the nearest and next nearest neighbor spin-spin correlations, obtained from MC simulations (see Fig. 4), bears a monotonous decrease with reducing temperature below room temperature. Given that the modes S1, S2 and S3 couple similarly to spin (see Table III) with the highest value of frequency in the NM state and the lowest one in stripe phase (AFM2), all the three Raman-active modes should soften as the temperature is lowered; the $T$-dependence of S2 mode is expected to be weaker because its coupling with spin is much smaller. This is consistent with our data in figure 3 where changes in the frequency below $T_{SM}$ are minimum for the S2 mode. These theoretical predictions are consistent with the observed softening of Raman modes at low temperature in our experiment. Secondly, the precise magnetic ordering in the low-temperature phase is mixed as reflected in low temperature values of the spin-spin correlation (Fig. 4), which should have been 0 and -1 for the first and second neighbor spin-correlations in the stripe phase. We suggest that the low temperature phase of CaFe$_2$As$_2$ is partially stripe anti-ferromagnetic and it is due to frustration coming from opposite signs of first and second neighbor exchange interactions between Fe. With decrease in temperature, we believe that the system undergoes a magnetic transition from paramagnetic to *short-ranged* AFM2 ~ 140 K [30] and softening of Raman active modes at the low temperatures is due to the spin-phonon coupling which remains strong as a function of $x$. This is understandable because such a coupling arises from the changes in the Fe-As-Fe bond angles (subsequently the super-exchange interactions) associated with atomic displacements (see Fig. 2) in S1 and S2 modes.

## 4. CONCLUSIONS

In conclusion, we have shown that all the three observed modes show anomalous temperature-dependence due to strong spin-phonon coupling in Ca(Fe$_{0.95}$Co$_{0.05}$)$_2$As$_2$.

Density functional calculations of phonons in different magnetic phases show strong spin-phonon coupling in parent and doped (122) system which is responsible for the observed softening of the phonon frequency with decreasing temperature. The anomalous hardening of one of the Raman active mode below the superconducting transition temperature is attributed to the coupling of the mode with the superconducting quasi-particle excitations, yielding an upper limit of $2\Delta/k_B T_c \sim 15$. Results obtained here suggest that the interplay between phonons and spin degrees of freedom are crucial to unravel the underlying physics responsible for the pairing mechanism in iron-pnictides.

## Acknowledgments

PK, AB and AK acknowledge CSIR, India, for research fellowship. AKS and UVW acknowledge DST, India and DAE Outstanding Researcher Fellowship for financial support. The authors at Dresden thank M. Deutschmann, S. Pichl, and J. Werner for technical support and their work was supported by the DFG program FOR 538 and BE1749/13 project.

Table-I: Energies of internally relaxed structures FM, AFM1 and AFM2 (stripe) magnetic ordering and corresponding stresses on unit cell.

| Magnetic order | Energy (2*eV/fmu) | $\sigma_{xx}$ (kB) | $\sigma_{zz}$ (kB) | $\sigma_{xy}$ (kB) |
|---|---|---|---|---|
| NM | -49.501 | -76 | -131 | 0 |
| FM | -54.051 | -35 | -21 | 0 |
| AFM1 | -53.722 | 9 | -17 | 0 |
| AFM2 (Stripe) | -54.487 | -12 | 9 | **20** |

Table-II: List of the experimental observed frequencies at 4K in $Ca(Fe_{0.95}Co_{0.05})_2As_2$ and calculated frequencies for $Ca(Fe_{1-x}Co_x)_2As_2$ using LDA+U( = 4eV for Fe and Co), obtained with Quantum Espresso based calculations. $\Delta\omega$ represent the difference between frequency in NM and AFM2 phase.

| Mode Assignment | Experimental $\omega$ (cm$^{-1}$) | Calculated $\omega$ (cm$^{-1}$) | | | | | |
|---|---|---|---|---|---|---|---|
| | | x = 0.0 | | | x = 0.25 | | |
| | | NM | AFM2 | $\Delta\omega/\omega_{nm}$ (%) | NM | AFM2 | $\Delta\omega/\omega_{nm}$ (%) |
| S1 $B_{1g}$ (Fe) | 205 | 211 | 171 | 18.9 | 215 | 160 | 25.5 |
| S2 $A_{1g}$ (As) | 215 | 227 | 208 | 8 | 239 | 216 | 9.6 |
| S3 $E_g$ (As and Fe) | 267 | 320 | 279 | 12.8 | 306 | 270 | 11.7 |

Table-III: Zone center phonons (frequencies given in cm$^{-1}$) for NM, FM, AFM1 and AFM2 ordering, giving an estimate of the spin-phonon coupling, obtained with VASP-based calculations.

| Mode | NM | FM | AFM1 | AFM2 |
|---|---|---|---|---|
| $B_{1g}$ (S1) | 204 | 180 | 150 | 197 |
| $A_{1g}$ (S2) | 207 | 185 | 203 | 201 |
| $E_g$ (S3) | 301 | 223 | 237 | 248 |
| $E_g$ | 156 | 109 | 115 | 89 |

**FIGURE CAPTION**

FIG.1. (Color online) Unpolarised Raman spectra of Ca(Fe$_{0.95}$Co$_{0.05}$)$_2$As$_2$ at 4 K. Solid (thin) lines are fit of individual modes and solid (thick) line shows the total fit to the experimental data (circle). Inset (a) shows the mode S3 at few temperatures.

FIG.2. (Color online) Eigen modes corresponding to Raman modes in Table ( III ).

FIG.3. (Color online) Temperature dependence of the modes S1, S2 and S3. Solid lines are linear fit in a given temperature window.

FIG.4. (Color online) Temperature dependence of the spin correlation function ( $<S_i.S_j>$ ) for CaFe$_2$As$_2$ obtained from Monte Carlo simulations. Solid lines are guide to the eye.

Figure1:

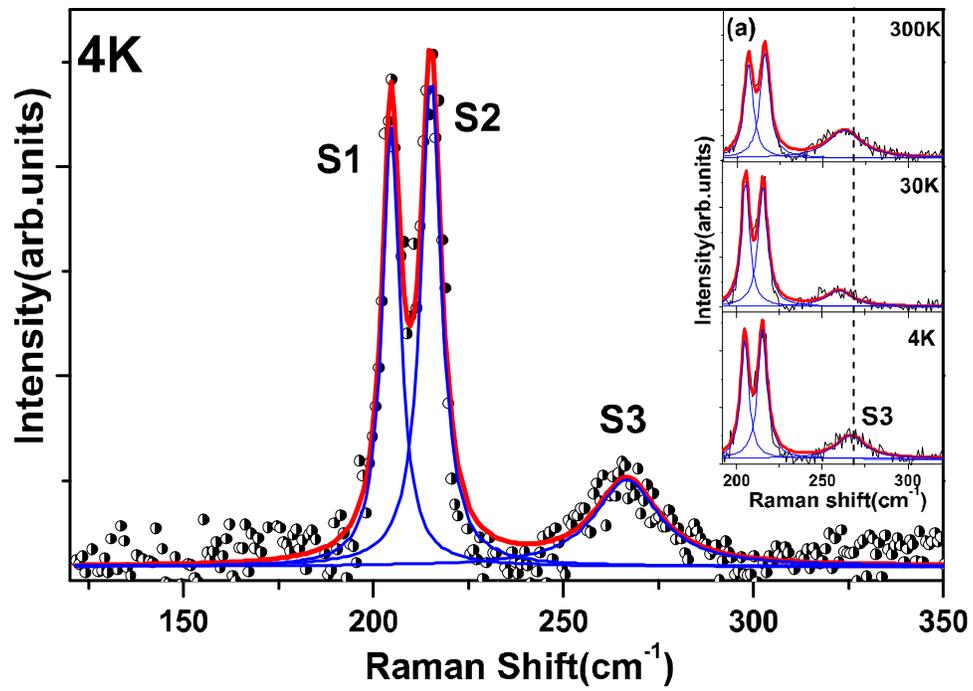

Figure2:

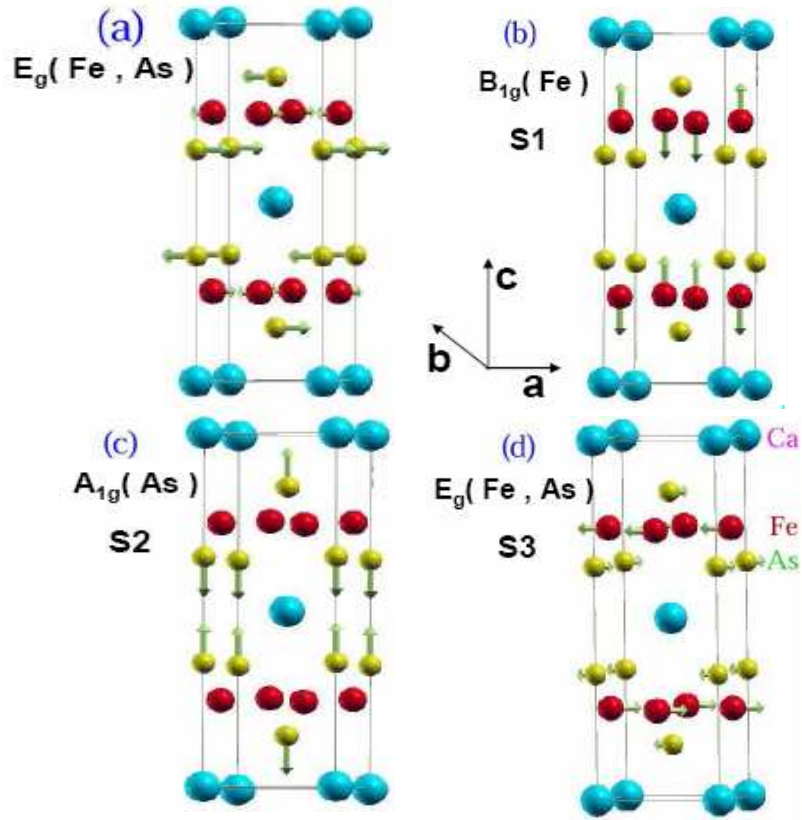

Figure3:

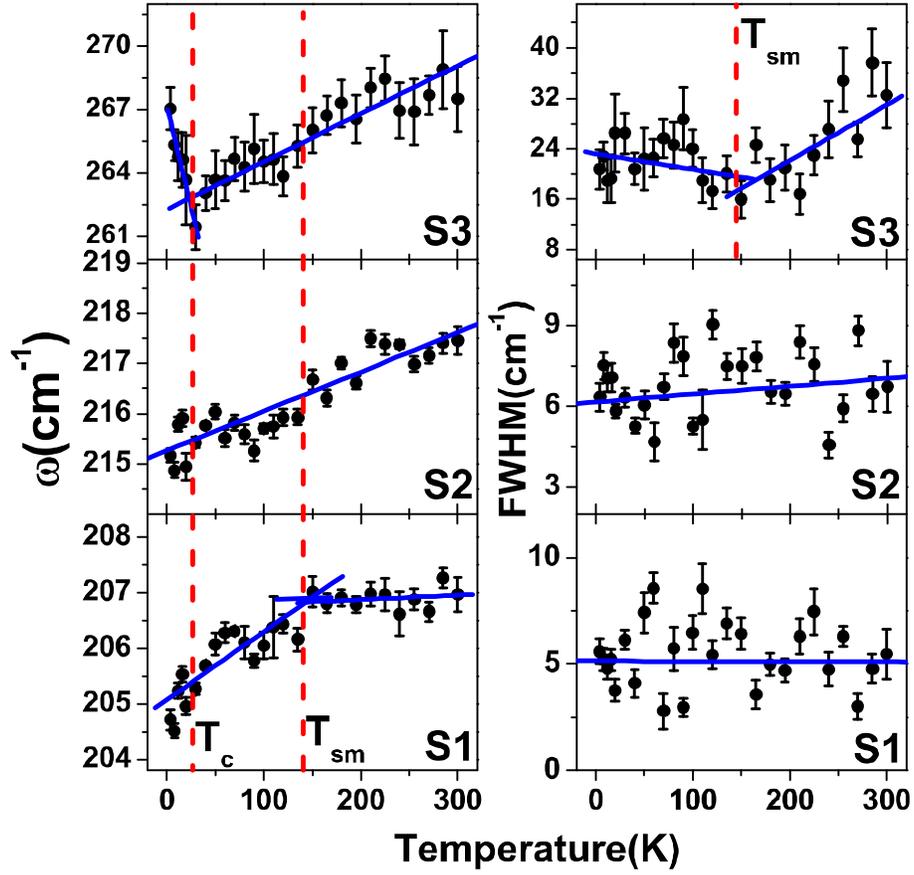

Figure 4:

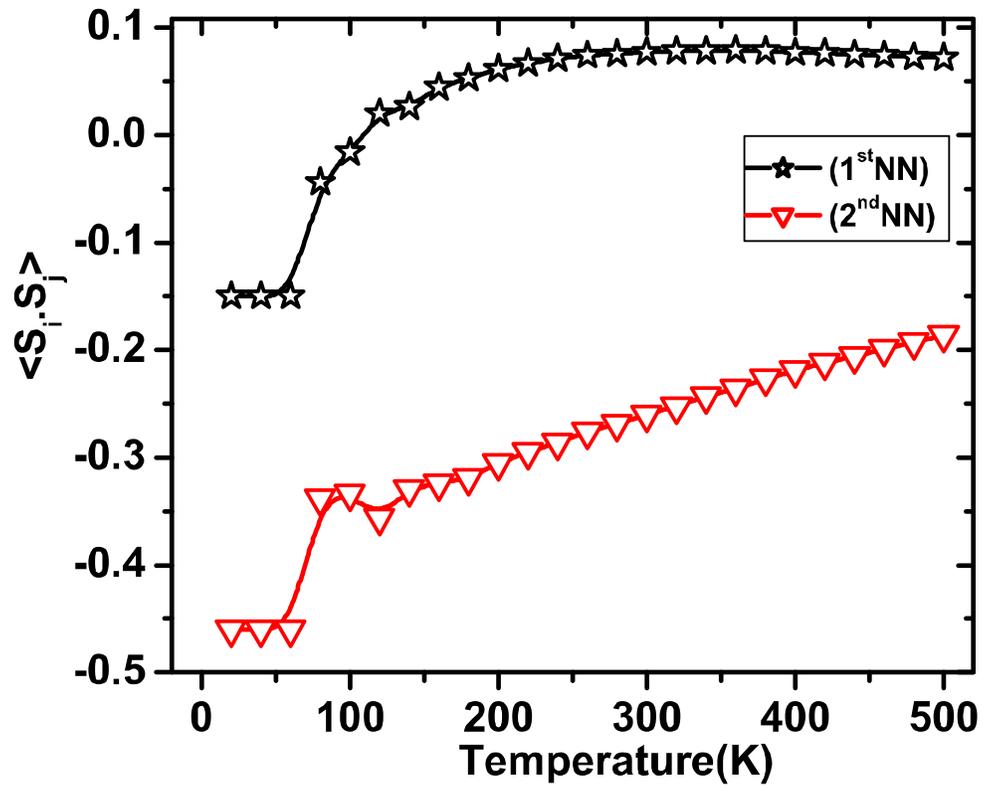